\documentclass[oldversion]{aa}
\usepackage{graphicx}
\usepackage{txfonts}
\usepackage{natbib}
\begin{document}

\title{OGLE-TR-211 -- a new transiting inflated hot Jupiter from the
\\ OGLE survey and ESO LP666 spectroscopic follow-up program
\thanks{Based on observations made with the FORS1 camera
and the FLAMES/UVES spectrograph at the VLT, ESO, Chile (programmes
07.C-0706, 076.C-0122 and 177.C-0666) and 1.3-m Warsaw Telescope at
Las Campanas Observatory, Chile.}}

\author{A. Udalski$^{1,2}$, F. Pont$^3$, D. Naef$\ ^4$, C. Melo$^4$,
F. Bouchy $^5$, N.C. Santos$^{6}$,  C. Moutou$^7$, R.F. D{\'i}az$^8$, 
W. Gieren$^{9}$, M. Gillon$^3$, S. Hoyer$^{10}$, M. Mayor$^3$, T. Mazeh$^{11}$,
D. Minniti$^{12}$, G. Pietrzy{\'n}ski$^{1,2,9}$, D. Queloz$^3$, S. Ramirez$^5$,
M.T. Ruiz$^{10}$, O. Tamuz$^{11}$, S. Udry$^3$, M.
Zoccali$^{12}$, M. Kubiak$^{1,2}$, M.K. Szyma{\'n}ski$^{1,2}$,
I. Soszy{\'n}ski$^{1,2}$, O. Szewczyk$^{1,2,9}$, K. Ulaczyk$^{1,2}$,
{\L}. Wyrzykowski$^{2,13}$}

\offprints{udalski@astrouw.edu.pl}

\institute{
$^1$ Warsaw University Observatory, Al. Ujazdowskie 4, 00-478, Warsaw, Poland\\
$^2$ The OGLE Team\\
$^3$  Observatoire de Gen\`eve, 51 Chemin des Maillettes,1290 Sauverny,
    Switzerland\\
$^4$ European Southern Observatory, Casilla 19001, Santiago 19, Chile\\
$^5$ Institut d'Astrophysique de Paris, 98bis Bd Arago, 75014 Paris, France\\
$^{6}$ Centro de Astrof\'\i sica, Universidade do Porto, Rua das Estrelas, 4150-762 Porto, Portugal\\
$^7$ Laboratoire d'Astrophysique de Marseille, Traverse du Siphon, BP8, Les Trois Lucs, 13376 Marseille
  cedex 12, France\\    
$^8$ Instituto de Astronomia y F\'\i sica del Espacio (CONICET-UBA), Buenos Aires, Argentina \\
$^{9}$ Departamento de Fisica, Astronomy Group, Universidad de Concepci\'on, Casilla 160-C, Concepci\'on, Chile\\
$^{10}$ Department of Astronomy, Universidad de Chile, Santiago, Chile\\
$^{11}$ School of Physics and Astronomy, R. and B. Sackler Faculty of Exact Sciences, Tel Aviv University, Tel Aviv, Israel\\   
$^{12}$ Departmento de Astronom\'ia y Astrof\'isica, Pontificia Universidad Cat\'olica de Chile, Casilla 306, Santiago 22, Chile\\
$^{13}$ Institute of Astronomy, University of Cambridge, Madingley Road,
Cambridge CB3 0HA, UK}

\date{Received date / accepted date}

\authorrunning{A. Udalski et al.}
\titlerunning{OGLE-TR-211 -- a new transiting hot Jupiter}

\abstract{
We present results of the photometric campaign for planetary and
low-luminosity object transits conducted by the OGLE survey in 2005 season
(Campaign \#5). About twenty most promising candidates discovered in these
data were subsequently verified spectroscopically with the VLT/FLAMES
spectrograph.

One of the candidates, OGLE-TR-211, reveals clear changes of radial
velocity with small amplitude of 82 m/sec, varying in phase with
photometric transit ephemeris. Thus, we confirm the planetary nature of
the OGLE-TR-211 system. Follow-up precise photometry of OGLE-TR-211 with
VLT/FORS together with radial velocity spectroscopy supplemented with 
high resolution, high S/N VLT/UVES spectra allowed us to derive
parameters of the planet and host star. OGLE-TR-211b is a hot Jupiter
orbiting a F7-8 spectral type dwarf star with the period of 3.68 days.
The mass of the planet is equal to $1.03\pm0.20$~M$_{\rm Jup}$ while its
radius 1.36 $^{+0.18}_{-0.09}$~R$_{\rm Jup}$. The radius is about 20\%
larger than the typical radius of hot Jupiters of similar mass.
OGLE-TR-211b is, then, another example of inflated hot Jupiters -- a
small group of seven exoplanets with large radii and unusually small
densities -- objects being a challenge to the current models of
exoplanets.}


\maketitle

\section{Introduction}
The discovery of the first transit of an extrasolar planet in HD209458
\citep{ch00,he00} opened a new era in the extrasolar planet field -- the
epoch of searches for extrasolar planets with large scale photometric
surveys. While this channel of finding exoplanets developed relatively
slowly in the first couple of years after that discovery, the last two
years brought a fast acceleration due to maturing of the search methods. At
present about 25 transiting planets are known (cf. {\it
http://obswww.unige.ch/~pont/TRANSITS.htm)} constituting about 10\% of all
known exoplanets.

Transiting planets play especially important role among the known
exoplanets. These are the only objects for which the most important
parameters like radius, mass and density can be precisely derived from
observations providing direct comparison with models and, thus, allowing
better understanding of the exoplanet structure and evolution. Also a
large variety of follow-up observations can be performed on transiting
system allowing, for example, studies of planetary atmospheres, search
for other planetary companions etc.

Transiting planets are discovered via two channels: photometric
follow-up of spectroscopically found exoplanets and classical transit
method approach -- large scale photometric surveys providing transiting
candidates that have to be then verified and confirmed
spectroscopically. Both approaches have been successful in providing
examples of very large diversity among transiting planets -- regular hot
Jupiters, very hot Jupiters on extremely short 1--2 day period orbits,
inflated objects with radii much larger than expected from modeling or
small Neptune-sized objects. Large number of new extrasolar transiting
planets are expected to be discovered in the next couple of years from the
space missions like Corot or Kepler, in particular small size planets
undetectable in ground-based searches.

The Optical Gravitational Lensing Experiment (OGLE) was the first
successful photometric survey that discovered large number of transiting
candidates that were subsequently verified spectroscopically
\citep{ud02a,ud02b,ud02c,ud03}. Five of these candidates, OGLE-TR-56,
OGLE-TR-113, OGLE-TR-132, OGLE-TR-111 and OGLE-TR-10 turned out to be
extrasolar planetary systems \citep{ko03,bo04,ko04,po04,bo05,ko05}, the
first extrasolar planets discovered with the transit method. Moreover,
two planet-sized stars of lowest known masses were also found
\citep{po05b,po06}. Beside that the OGLE photometric transit campaigns 
provided a huge observational material allowing better understanding the
problems of photometric transit searches, data systematics etc.
\citep{go06,pz06}.

OGLE transit campaigns have been conducted every year since 2001.
Results of the spectroscopic follow-up of candidates discovered during
the Campaign \#1 (2001) and \#2 (2002) were published by \citet{bo05}
and \citet{po05a}, respectively. Recently, analysis of the near
threshold candidates from the Campaign \#2 led to the discovery of the
sixth OGLE transiting exoplanet, OGLR-TR-182 \citep{po07}.

In this paper we present results of the OGLE Campaign \#5 conducted in 2005
and spectroscopic follow-up of the best transiting candidates detected in
these data. A new transiting extrasolar planetary system OGLE-TR-211 was
found among them. In the following Sections we provide the details of the
photometric and spectroscopic observations of OGLE new candidates and
derive parameters of a planetary companion in OGLE-TR-211 system.

\section{OGLE planetary transit campaign \#5}

OGLE planetary transit campaigns became a standard part of the OGLE
observing schedule. About 75\% of observing time is devoted to this
sub-project of the OGLE survey every southern fall -- from February to
April. The OGLE planetary campaign \#5 was conducted by OGLE in the 2005
observing season from February 2, 2005 to June 23, 2005. Observations
were carried out with the 1.3-m Warsaw Telescope at Las Campanas
Observatory, equipped with $8192\times8192$ pixel CCD mosaic camera
\citep{uda3}. The observing strategy was similar to that of the previous
campaigns, in particular \#3 and \#4 \citep{ud04}. Because the
experience gathered during the spectroscopic observations of the first
OGLE candidates indicated troubles in reasonable spectroscopic follow-up
of the faintest objects from typical OGLE candidate lists, the exposure
time during Campaign \#5 was shortened to 120 seconds. This allowed to
add one more field increasing the number of observed fields to four and
covering about 1.4 square degrees of the Galactic disk regions in the
Carina constellation. The cadence time for each of these four fields was
about 16 minutes and their equatorial coordinates are listed in Table~1.
All observations were collected through the {\it I}-band filter. The
median seeing of the all observations was about 1.1 arcsec. Altogether
about 1320 images of each of the fields were secured during the
Campaign. Also a few {\it V}-band observations were taken for color
information and CMD construction.

\begin{table}
\caption{Equatorial coordinates of the fields observed during the OGLE 
Campaign \#5.}
\label{table:1}
\centering
\begin{tabular}{ccc}
\hline
\hline
\noalign{\vskip 3pt}
Field  & RA(J2000) & DEC(J2000)  \\
\hline
\noalign{\vskip 3pt}
CAR107 & 10:47:15 & $-62{:}00{:}25$\\
CAR108 & 10:47:15 & $-61{:}24{:}35$\\
CAR109 & 10:42:10 & $-62{:}10{:}25$\\
CAR110 & 10:42:15 & $-61{:}34{:}35$\\
\hline
\end{tabular}
\end{table}

Collected data were reduced in similar way as those of Campaigns \#3 and
\#4 \citep{ud04}. The selection of transit candidates was also performed
similarly: all non-variable objects, that is those with the {\it rms} of
the average magnitude of all observations smaller than 0.015 mag, were
subject to detrending algorithm of \citet{kr03} and then transit search
procedure using the BLS algorithm of \citet{kv02}. Altogether light
curves of about 50000 stars were analyzed. The list of transiting
candidates was prepared after careful visual inspection of candidates
that passed the BLS criteria. Clear false cases triggered, for example,
by noisy light curves were removed, as well as objects with the depth of
transits larger than 50 mmag, those with clear signature of ellipsoidal
effect indicating massive secondary, those with clear V-shape of
transits indicating grazing eclipses of a binary star and those with the
number of individual transits smaller than three. For all the candidates
the limits on the size of a transiting companion were calculated as in
\citet{ud04}. All candidates with the companion having the lower limit
of radius larger than 0.2~R$_\odot$ were removed from the final list.

Table~2 lists the transit candidates from the OGLE Campaign \#5 that
passed the photometric search criteria. The naming convention follows
the standard OGLE convention, i.e., OGLE-TR-NNN. About twenty candidates
for transiting planetary systems were found. These objects were selected
as targets for spectroscopic follow-up observations conducted under the
LP666 Program. Photometry of all selected candidates is available from
the OGLE Internet archive at {\it
ftp://ftp.astrouw.edu.pl/ogle/ogle3/transits/tr201-219/}

\begin{table*}[ht]
\caption{Planetary and low-mass object transit candidates
from the OGLE Campaign \#5 conducted in 2005.}
\centering
\begin{tabular}{c@{\hspace{15pt}}
                c@{\hspace{15pt}}
                c@{\hspace{15pt}}
                c@{\hspace{15pt}}
                c@{\hspace{15pt}}
                c@{\hspace{15pt}}
                c@{\hspace{15pt}}
                l}
\hline
\hline
\noalign{\vskip 3pt}
Name & RA & DEC & Period & $T_0$ & $I$ & Depth & Status 	\\
     &  [2000] & [2000] & [days] & [Hel.JD] & [mag] &  [mag] &        \\ 
\hline
\noalign{\vskip 3pt}
OGLE-TR-201 &  10:48:31.61 &  $-62{:}01{:}01{.}8$ & 2.3680 & 2453404.860 & 15.6 & 0.016 & fast rotator  \\
OGLE-TR-202 &  10:46:06.06 &  $-61{:}52{:}11{.}3$ & 1.6545 & 2453404.438 & 13.6 & 0.017 & not observed  \\ 
OGLE-TR-203 &  10:49:32.04 &  $-61{:}35{:}38{.}2$ & 3.3456 & 2453406.178 & 15.6 & 0.014 & not observed  \\
OGLE-TR-204 &  10:47:39.44 &  $-61{:}19{:}00{.}8$ & 3.1097 & 2453405.515 & 14.8 & 0.026 & SB2           \\
OGLE-TR-205 &  10:45:53.40 &  $-61{:}09{:}22{.}1$ & 1.7501 & 2453404.601 & 16.0 & 0.015 & not observed  \\
OGLE-TR-206 &  10:45:22.00 &  $-61{:}28{:}28{.}8$ & 3.2658 & 2453403.246 & 13.8 & 0.006 & no variation  \\
OGLE-TR-207 &  10:40:10.41 &  $-61{:}53{:}55{.}5$ & 4.8170 & 2453406.978 & 14.3 & 0.021 & SB2           \\
OGLE-TR-208 &  10:39:54.18 &  $-61{:}58{:}07{.}7$ & 4.5025 & 2453406.567 & 15.3 & 0.022 & SB2           \\
OGLE-TR-209 &  10:40:56.27 &  $-62{:}14{:}20{.}2$ & 2.2056 & 2453403.657 & 15.0 & 0.022 & no variation  \\
OGLE-TR-210 &  10:40:44.21 &  $-62{:}13{:}21{.}6$ & 2.2427 & 2453403.012 & 15.2 & 0.032 & fast rotator  \\
OGLE-TR-211 &  10:40:14.51 &  $-62{:}27{:}19{.}8$ & 3.6772 & 2453406.271 & 14.3 & 0.008 & planet        \\
OGLE-TR-212 &  10:43:37.95 &  $-61{:}45{:}01{.}5$ & 2.2234 & 2453404.170 & 16.3 & 0.016 & blend?        \\
OGLE-TR-213 &  10:43:00.85 &  $-61{:}51{:}10{.}2$ & 6.5746 & 2453403.301 & 15.3 & 0.036 & SB2           \\
OGLE-TR-214 &  10:44:27.96 &  $-61{:}35{:}35{.}7$ & 3.6010 & 2453403.090 & 16.5 & 0.023 & SB1           \\
OGLE-TR-215 &  10:43:48.69 &  $-61{:}40{:}00{.}3$ & 4.9237 & 2453407.616 & 14.8 & 0.016 & no variation  \\
OGLE-TR-216 &  10:42:02.43 &  $-61{:}20{:}16{.}0$ & 1.9763 & 2453403.064 & 14.6 & 0.011 & blend?        \\
OGLE-TR-217 &  10:40:59.73 &  $-61{:}30{:}38{.}4$ & 5.7208 & 2453404.104 & 16.1 & 0.037 & no ccf        \\
OGLE-TR-218 &  10:41:12.74 &  $-61{:}28{:}20{.}8$ & 2.2488 & 2453404.697 & 14.5 & 0.020 & fast rotator  \\
OGLE-TR-219 &  10:40:53.40 &  $-61{:}43{:}15{.}1$ & 9.7466 & 2453405.683 & 15.1 & 0.032 & SB2           \\
\hline
\end{tabular}
\end{table*}

\section {Spectroscopic follow-up}

Spectroscopic follow-up observations of the OGLE Campaign \#5 transit
candidates were carried out during three observing slots allocated to the
LP666 program in April/May 2006, February 2007 and April 2007 on VLT
with the FLAMES spectrograph. Also, part of the Geneva group observing time
under program 07.C-0706, on the same instrument in February 2006 was
used. Unfortunately the weather conditions during all these observing runs
were exceptionally unfavorable with many cloudy or large seeing nights
($>2$~arcsec) what considerably limited the results. In particular after
the initial screening of all candidates only the most promising objects
were further observed. The remaining objects, contrary to the previous
follow-up observations \citep{bo04,po05a}, were left without full further
characterization. The strategy of the spectroscopic follow-up was identical
as described by \citet{po07}.

Table~2 includes the column with spectroscopic status resulting from the
collected spectra. Only one candidate turned out to be a very promising
planetary candidate revealing low amplitude radial velocity variation in
phase with photometric ephemeris, namely OGLE-TR-211. It was then
extensively observed during all following spectroscopic runs what allowed
to confirm its planetary nature. Three more candidates do not show
significant radial velocity variations. Nevertheless, the confirmation
of their planetary nature seems to be impossible as the lack of
variation does not necessarily proves the presence of a planetary
companion. Table~3 lists these objects and provides the upper limits on
radial velocity semi-amplitudes, $K$, at the 2-sigma level and the
corresponding maximum mass of a potential planet for a 1~M$_\odot$ host
star.

\begin{table}
\caption{OGLE transit candidates with no significant radial velocity variation.}
\centering
\begin{tabular}{ccc}
\hline
\hline
\noalign{\vskip 3pt}
Object  & $K$ limit & Mass limit  \\
        & [m/sec]   & [M$_{\rm Jup}$]\\
\hline
\noalign{\vskip 3pt}
OGLE-TR-206 & $<59$ & $<0.42$ \\
OGLE-TR-209 & $<16$ & $<0.16$ \\
OGLE-TR-215 & $<63$ & $<0.52$ \\
\hline
\end{tabular}
\end{table}

\section{OGLE-TR-211 candidate}

OGLE-TR-211 turned out to be the most promising candidate for a
transiting planetary system from the OGLE Campaign \#5 sample. After the
initial spectroscopic follow-up observations confirming this status it
was decided to allocate considerable amount of observing time for
precise characterization of this object. OGLE-TR-211 is relatively
bright (${\it I}\approx14.3$~mag) in the OGLE sample. Its equatorial
coordinates are listed in Table~2 while Figure~1 shows the finding chart
-- $2'\times2'$ OGLE {\it I}-band image.

\begin{figure}
\includegraphics[width=8.7cm, angle=90]{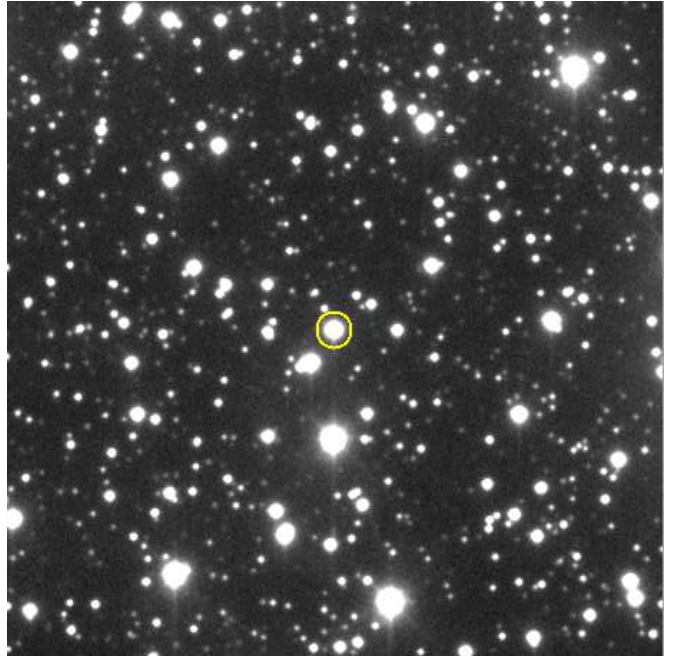}
\caption{Finding chart for OGLE-TR-211. Size of the {\it I}-band
OGLE subframe is $120"\times120"$, North is up and East to the left.}
\end{figure}

\subsection{Photometry}

The original OGLE 2005 season photometric data covered eight partial
transits. When folded with the period of 3.677 days they revealed a
clear transit about 8 mmag deep. The estimated lower limit of the
transiting companion radius was 0.12 R$_\odot$ making OGLE-TR-211 a very
good candidate for a transiting planet.

Due to limited, $\approx$3 months long, duration of the OGLE Campaign
the photometric period has limited accuracy of the order of
$\approx10^{-4}P$ or even worse in the case of low depth transits. To
confirm presence of transits in OGLE-TR-211 and refine the ephemeris
OGLE observed the star in the next observing seasons: 2006 and 2007. The
first additional transit data points were collected yet in January 2006
providing a necessary confirmation and update of the photometric orbit
for the coming spectroscopic follow-up observations. Up to June 2007
eight additional transits of OGLE-TR-211 were covered. Thanks to over
two years long baseline these observations allowed us to derive precise
photometric ephemeris of OGLE-TR-211, listed in Table~5.

Very small depth of transits combined with $\approx$5 mmag accuracy of
individual OGLE measurements makes the determination of precise shape of
transit difficult. Therefore, similarly to other cases of OGLE transiting
planets, it was decided to conduct a photometric follow-up of this object
on much larger telescope. Three photometric runs on VLT with the FORS
camera were performed in the period of March/June 2006.

The first run VLT images of OGLE-TR-211 were obtained on March 16, 2006
through the {\it V}-band filter with the exposure time of 8 seconds. They
cover the pre-ingress, ingress and bottom of the transit. Unfortunately the
egress of the transit was missed. Two additional runs were carried out on
May 25, 2006 and June 30, 2006. In these cases observations were obtained
through the {\it V} and {\it R} filters and exposure time varied from 25 to
80 seconds depending on seeing. Unfortunately, both these runs cover
practically non-variable part of the transit light curve: the May
observations -- flat bottom of the transit with only a small trace of
starting egress at the very end, the June observations -- only post transit
light curve with the very end of egress at the beginning. Thus, the full
reconstruction of the precise VLT light curve OGLE-TR-211 became difficult.

\begin{figure}
\includegraphics[width=8.7cm, bb=15 45 515 340]{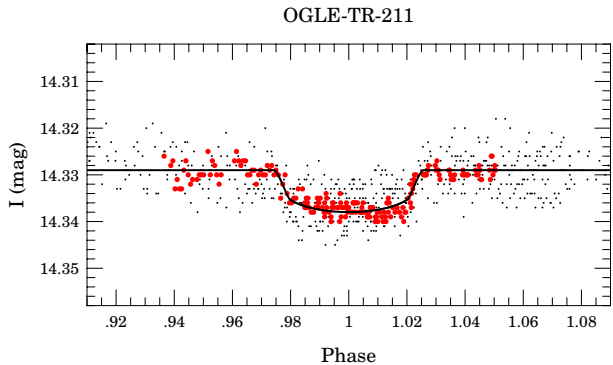}
\caption{Light curve of OGLE-TR-211. Small dots indicate OGLE data points
while large dots VLT observations. Continuous line represents the best
model of the transit.}
\end{figure}

The VLT data were reduced using the OGLE data pipeline \citep{uda3}, based
on the Difference Image Analysis (DIA) method. Because the March run that
defines well considerable part of the transit shape was obtained in the
{\it V}-band, only {\it V}-band images from the remaining runs were
used. To have similar time resolution and lower the scatter of the March
observations these data were binned to have an effective resolution of 3
minutes. The March light curve indicates that the transit depth is about 8
mmag -- practically identical with that resulting from OGLE observations
what is reassuring. Therefore the May data were adjusted in such a way that
the observed flat bottom of the transit corresponds to the level of the
bottom observed in March. Similarly, the June photometry was shifted so the
post-egress part of transit corresponds to the pre-ingress part observed in
March. Finally, the VLT photometry was folded with the photometric period
resulting from OGLE observations. Figure~2 shows the reconstructed VLT
light curve of OGLE-TR-211 plotted using large dots and overimposed on the
OGLE light curve (small dots).

The VLT data were also reduced using the DecPhot PSF fitting software
\citep{gi06}. While the results were generally similar, due to 
difficulties with the normalization between runs this photometry was not
used for further analysis of OGLE-TR-211.

\subsection{Radial velocity follow-up}

Twenty high-resolution spectroscopic observations of OGLE-TR-211 were
obtained with the FLAMES/UVES spectrograph on VLT during four
spectroscopic follow-up runs. Strategy of observations and reduction
procedures were identical as in \citet{po07}. The resulting radial
velocities of OGLE-TR-211 are listed in Table~4. Similarly as in
\citet{po07} we added 45 m/sec error in quadrature to the photon noise
radial velocity errors to account for possible systematic errors.

After two 2006 runs it was clearly found that the radial velocities of
OGLE-TR-211 follow the photometric period and reveal $\approx100$ m/sec
semi-amplitude variation in appropriate phase with planetary
interpretation. However, during the last 2007 April run, it was noted
that while this variation is clearly seen, the zero point of radial
velocities shifted by about 180 m/sec. This is considerably larger shift
than the shifts noted by \citet{po07} during analysis of the recently
discovered transiting planet OGLE-TR-182.

Although it cannot be fully ruled out that the shift has some
instrumental origin it is more likely that it corresponds to the real
change of the average radial velocity. Unfortunately, the collected
dataset is too limited to draw any sound conclusions about its origin.
We can only speculate that it may be caused by the presence of additional
companion in the OGLE-TR-211 system having wide orbit with a period of
several years. Further additional spectroscopic observations are
necessary to verify this interpretation.

\begin{table}
\caption{Radial velocity measurements for OGLE-TR-211.}
\begin{tabular}{ccccc}
\hline
\hline
\noalign{\vskip 3pt}
Date & Phase & VR & VR$_{\rm rectified}$ &  $\sigma_{VR}$   \\
$$[JD-2450000] & & [km s$^{-1}$] & [km s$^{-1}$]  & [km s$^{-1}$]\\
\hline
\noalign{\vskip 3pt}
3793.81192 &   0.3889 &   18.784  &  18.784 &    0.055 \\
3794.75446 &   0.6452 &   18.866  &  18.866 &    0.053 \\
3852.50001 &   0.3487 &   18.720  &  18.720 &    0.053 \\
3853.51724 &   0.6254 &   18.875  &  18.875 &    0.053 \\
3854.54151 &   0.9039 &   18.841  &  18.841 &    0.051 \\
3855.50251 &   0.1652 &   18.711  &  18.711 &    0.055 \\
3855.59958 &   0.1916 &   18.816  &  18.816 &    0.060 \\
3858.49787 &   0.9798 &   18.906  &  18.906 &    0.054 \\
3881.57542 &   0.2556 &   18.766  &  18.766 &    0.053 \\
4143.83440 &   0.5753 &   18.860  &  18.860 &    0.046 \\
4144.71247 &   0.8141 &   18.911  &  18.911 &    0.046 \\
4145.72709 &   0.0900 &   18.762  &  18.762 &    0.048 \\
4149.81577 &   0.2019 &   18.681  &  18.681 &    0.047 \\
4150.77874 &   0.4638 &   18.891  &  18.891 &    0.046 \\
4203.55370 &   0.8156 &   18.745  &  18.925 &    0.046 \\
4204.51070 &   0.0759 &   18.623  &  18.803 &    0.046 \\
4205.59031 &   0.3695 &   18.657  &  18.837 &    0.046 \\
4207.70041 &   0.9433 &   18.658  &  18.838 &    0.046 \\
4208.68020 &   0.2097 &   18.578  &  18.758 &    0.046 \\
4209.61470 &   0.4639 &   18.517  &  18.697 &    0.049 \\
\hline
\end{tabular}
\end{table}

For the purpose of characterization of the planetary companion of
OGLE-TR-211 the mean shifts of each observing run compared to the first
one were fitted. Only the shift of the April 2007 run turned out to be
statistically significant, It was subtracted from original radial
velocities. Rectified radial velocities of OGLE-TR-211 are also listed
in Table~4.

\begin{figure}
\includegraphics[width=8.7cm, bb=15 45 510 450]{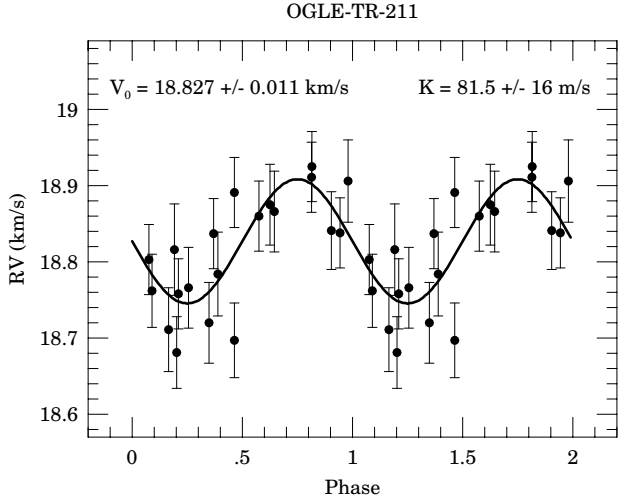}
\caption{Radial velocity observations for OGLE-TR-211, phased with the
photometric transit ephemeris. Solid line represents the best fit
assuming no eccentricity. Two periods are presented for clarity.}
\end{figure}

Because blending of an eclipsing star with a bright third light source
can cause radial velocity variation comparable with planetary signal,
line profiles of the spectra of OGLE-TR-211 were examined for
correlation of the average line bisectors with phase and observed radial
velocities \citep[e.g.,][]{sa02,to04}. No such correlation was found
allowing to rule out blending scenario.

\subsection{Host star spectroscopy}

To derive parameters of the host star of OGLE-TR-211 high resolution
spectra were obtained with the UVES spectrograph on VLT working in the
slit mode on May 7 and 8, 2007. The spectra were summed so the coadded
spectrum reached a S/N of about 100 in the $\lambda\approx6707$~\AA~
Lithium line region. Measurements were done following the method
described in \citet{sa06}. The results are listed in Table~5.
Temperature and gravity indicates that the host star is a dwarf of F7-8
spectral type and metallicity somewhat larger than solar.

\subsection{OGLE-TR-211 parameters}

The parameters of OGLE-TR-211 were derived with the standard method used
in the case of the previous OGLE planets \citep{bo04,po07}. The
photometry with red noise value equal to the photon noise was fitted for
$P$, $T$, $b$, $R/R_*$ and $a/R_*$. Eccentricity of the orbit in the
final fitting was assumed to be zero, as practically all the observed
orbits of short period transiting planets are circularized by tidal
interactions. Also our preliminary fit of the radial velocity curve with
the orbital period fixed at the photometric value yielded eccentricity
$e = 0.16\pm0.22$, consistent with circular orbit at better than
$1\sigma$ level.

Then the radius and mass of the host star were estimated with the
maximum likelihood method combining the observed spectroscopic
parameters of the host star and \citet{gr02} stellar evolution models.
Table~5 lists the parameters of the host star and transiting planet in
the OGLE-TR-211 system.

It has to be noted that the transit shape is not precise enough to
determine the inclination angle very precisely. Nevertheless, the size
of the planet is relatively well constrained thanks to the accuracy of
the stellar spectroscopic parameters.

The O--C weighted {\it rms} scatter around the final photometric model
of the OGLE-TR-211 system is equal to $\sigma=1.6$~mmag and
$\sigma=4.1$~mmag for the VLT and OGLE dataset, respectively. The
weighted {\it rms} of radial velocity observations relatively to the
best fit is $\sigma_{\rm RV}=48$~m/s.

\begin{table}[th] 
\caption{Parameters of the OGLE-TR-211 system.}
\begin{tabular}{ll}
\hline
\hline
\noalign{\vskip 3pt}
Period       [days]      & $3.67724\pm0.00003$\\
Transit epoch [Hel. JD]  & $2453428.334\pm0.003$\\
VR semi-amplitude  [m/s] & $82\pm16$\\
Semi-major axis [AU]     & $0.051\pm0.001$\\
Radius ratio             & $0.085\pm0.004$\\
Orbital angle [$^\circ$] & $>82.7$\\
                         & \\
Light Curve Model:       & \\
OGLE {\it rms} [mmag]    & 4.1 \\
VLT  {\it rms} [mmag]    & 1.6 \\
                         & \\
Radial Velocity Model:   & \\
{\it rms} [m/s]          & 48 \\
                         & \\
$T_{\rm eff}$ [K]        & $6325\pm91$\\
$\log g$                 & $4.22\pm0.17$\\
$\eta$ [km s$^{-1}$]     & $1.63\pm0.21$\\
$$[Fe/H]                 & $0.11\pm0.10$\\ 
                         & \\
Star radius  [R$_\odot$] & $1.64^{+0.21}_{-0.07}$\\
Star mass [M$_\odot$]    & $1.33\pm0.05$\\
                         & \\
Planet radius [R$_{\rm Jup}$]& $1.36^{+0.18}_{-0.09}$\\
Planet mass [M$_{\rm Jup}$]      & $1.03\pm0.20$\\ 
\hline
\end{tabular}
\end{table}

\section{Discussion}

Parameters of the transiting companion of OGLE-TR-211 host star
indicate that it is a planetary object belonging to the hot Jupiter
class of extrasolar planets. It brings the total number of extrasolar
planets discovered by the OGLE survey to seven -- significantly lowering
the discrepancy between the number of typical hot Jupiters and shortest
period very hot Jupiters in the OGLE sample of extrasolar transiting
planets.

OGLE-TR-211b orbits an F7-8 main sequence star. The system parameters
closely resemble that of the recently discovered system HAT-P-6
\citep{no07} with the only difference that the host star in OGLE-TR-211 is
more metal abundant by $\approx0.2$~dex. The position of the planet on the
mass-orbital period diagram is similar to the position of other transiting
exoplanets of similar mass and period supporting the relation noted first
by \citet{ma05}.

While the mass of a new hot Jupiter OGLE-TR-211b is very similar to the
mass of another hot Jupiter OGLE-TR-182b recently found by \citet{po07}
its radius is about 20\% larger. The mean density of OGLE-TR-211b is
about 0.5 g/cm$^3$, placing it among a group of seven objects with the
lowest densities: Tres-4, HD209458, HAT-P-1, WASP-1, HAT-P-4 and HAT-P-6
\citep{kv07}. All these objects have inflated radii compared to other
transiting hot Jupiters of similar mass. Such planets pose a problem to
the planetary models and are crucial objects for testing the proposed
scenarios explaining ``bloated'' hot Jupiters \citep{bu07,gu06}. It is
worth noting that the metallicity of the OGLE-TR-211 host star falls in
the middle of the range of metallicities of ``bloated'' Jupiters host
stars (recently increased significantly by the HAT-P-6 system)
suggesting that metallicity cannot be the only parameter responsible for
inflating the radii of hot Jupiters.

\begin{figure}
\includegraphics[width=9cm, bb=75 392  505 660]{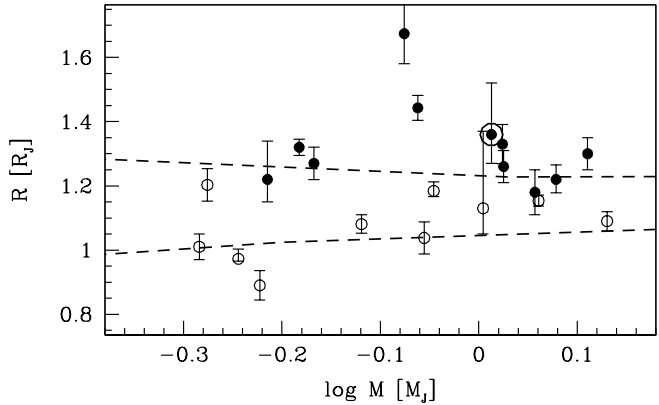}

\caption{Mass-radius diagram for transiting hot Jupiters. Closed symbols
indicate host stars with $T_{\rm eq}>1500$ K. OGLE-TR-211 is   indicated
by the circled dot. The lines show the models of \citet{gu05} in two
extreme cases:  cold gas planet with a $15 M_\oplus$ core  and   $T_{\rm
eq}=2000$~K planet without core. The five large planets with   radius
error bars above the upper models are all strongly irradiated.}
\end{figure}

On the other hand, the characteristics of the OGLE-TR-211 system  
reinforce the association of ``bloated'' close-in giant planets with  
strong irradiation from the host star. Figure~4 shows the mass-radius
diagram for the presently known transiting planets, classified  
according to their equilibrium temperature. \footnote{$T_{\rm eq} =
T_*((1-A)/4\,F)^{1/4}(R_*/a)^{1/2}$ using for all planets
a Bond albedo $A=0.3$ and a heat redistribution factor $F=1/2$.} 

All anomalously large planets have equilibrium temperatures hotter than  
1500~K.  This would require an increasingly unlikely coincidence if  
the anomalously large radii were explained by dynamical effects such  
as obliquity tide or eccentricity pumping by an unseen companion  
\citep{wi05}. It rather tends to favor  explanations directly linking
the excess of radius with the incident flux like such as  proposed by
\citet{gu06} (that a fraction of  the incident flux energy is converted
to mechanical energy by a yet undetermined physical process) -- over
explanations with a more  indirect causal relation, such as that put
forward by \citet{bu07} and \citet{ch07} invoking increased opacity
or internal density gradients.

As we already noted in Section~4.2 the mean level of radial velocity
variation of the OGLE-TR-211 host star was shifted by about 180 m/sec
during the last April 2007 run. Although the real nature of this shift
is unknown at this stage the tempting explanation is that it is caused
by the presence of additional companion, possibly of planetary or
stellar nature, in the OGLE-TR-211 system. If confirmed by  additional
follow-up observations, OGLE-TR-211 would be the first multi-planetary
system with a transiting planet or the first moderately-wide binary
system hosting a transiting hot Jupiter.

Also, further additional high accuracy photometric follow-up
observations of the OGLE-TR-211 transits are necessary for better
constraining the inclination and radius of the planet. Although
OGLE-TR-211, similarly to other OGLE transiting planets, is too faint
for the currently available facilities to carry out some follow-up
observations like IR photometry, the precise long term transit timing
allowing search for another planetary objects in the OGLE-TR-211 system
is feasible.

\begin{acknowledgements}

The OGLE project is partially supported by the Polish MNiSW grant
N20303032/4275. WG, DM, GP, MTR and MZ gratefully acknowledge support
for this work from the chilean FONDAP Center of Astrophysics 15010003.
NCS acknowledges the support from Funda\c{c}\~ao para a Ci\^encia e a
Tecnologia, Portugal, in the form of a grant (reference
POCI/CTE-AST/56453/2004), and support by the EC's FP6 and by FCT (with
POCI2010 and FEDER funds), within the HELAS international collaboration.
 
\end{acknowledgements}

\end{document}